\documentclass[prl,showpacs,preprintnumbers,twocolumn]{revtex4}
\usepackage{graphicx}

\begin{document}

\title{Ising lines: natural topological defects within chiral ferroelectric domain walls}

\author{ V. Stepkova$^{\rm a}$}
\author{ P. Marton$^{\rm a}$}
\author{ J. Hlinka$^{\rm a}$}
\email{hlinka@fzu.cz}
\affiliation{$^{\rm a}${\em{Institute of Physics, The Czech Academy of Sciences, Na Slovance 2, 18221 Praha 8, Czech Republic}}}
\date{\today}

\begin{abstract}
Phase-field simulations demonstrate that the polarization order-parameter field in the Ginzburg-Landau-Devonshire model of rhombohedral ferroelectric BaTiO$_3$ allows for an interesting linear defect, stable under simple periodic boundary conditions. This linear defect, termed here as Ising line,  can be described as about 2\,nm thick intrinsic paraelectric nanorod acting as a highly mobile borderline between finite portions of Bloch-like domain walls of the opposite helicity.
These Ising lines play the role of  domain boundaries associated with the Ising-to-Bloch domain wall phase transition.
\end{abstract}




\maketitle


Perovskite ferroelectrics are the key materials in the current ferroelectric research and a considerable effort was recently devoted to the investigations of their domain structure and their domain wall properties. Much less is known about ferroelectric line defects, although this topic is also gaining attention \cite{skyrme,GatalanSeidelRamesh2012,MorozovskaPRB2006,
ProsandeevPRB2007,NaumovPRL2008,Slutsker2008,Gru08,Bal09,Sta11,Baudry,Louis}, for example, in relation to magnetic vortices and skyrmions \cite{skyrme,GatalanSeidelRamesh2012}.
Ferroelecric domains and line defects may be inspiring in other research areas, too, for example in the particle physics \cite{zurek}.

\begin{center}
\begin{figure}
\includegraphics[width=7.5cm]{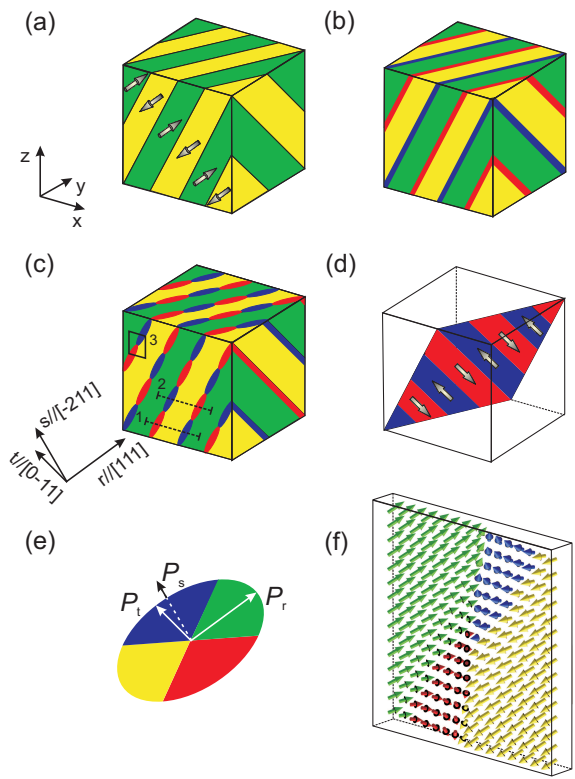}
\caption{
Predicted domain structure with Ising lines.
 Panel (a) shows schematically a domain structure  with $\langle\bar{2}11\rangle$ oriented domain walls between $[111]$ and $[\bar{1}\bar{1}\bar{1}]$ oriented ferroelectric domains. In case of the rhombohedral phase of BaTiO$_3$,  these domain walls become polar with the inner polarization pointing around $[01\bar{1}]$ or $[0\bar{1}1]$  directions. When the inner polarization of the wall is alternating from the wall to wall as in panel (b), the structure is chiral. Panel (c) shows a racemic domain structure with the  investigated Ising lines, forming the borderlines between areas of the opposite inner polarization. Orientation of the inner polarization within selected domain wall is sketched in panel (d). Panels (e) and (f) show the symmetry adapted polarization coordinates and the core structure of the Ising lines, respectively.
} \label{fig1}
\end{figure}
\end{center}

One of the most intriguing recent result in this field is the prediction of so-called Bloch-like domain walls in BaTiO$_3$ and PbTiO$_3$, based on Ginzburg-Landau-Devonshire model \cite{Mart10,BlochPT,StepJPCM,EliseevYudin} and first-principles calculations
\cite{TahePRB2012, iniguez}.
Previously, we and others studied $[\bar{2}11]$-oriented 180-degree domain walls in the rhombohedral  BaTiO$_3$,  separating domains with spontaneous polarization parallel and antiparallel to $[111]$ (as usual we refer to the cubic axes of the parent paraelectric phase of BaTiO$_3$, see Fig.\,\ref{fig1}\,a). Calculations carried out within the Ginzburg-Landau-Devonshire model of Refs.\,\onlinecite{Hlin06,Ondrejkovic2009} suggest that this domain wall has a Bloch-like (bistable and chiral) structure \cite{Mart10, EliseevYudin}. At the same time, it has been shown that this ($[\bar{2}11]$-oriented) Bloch-like wall can be transformed to an achiral, Ising-like domain wall by a moderate uniaxial stress \cite{StepJPCM}. The transformation proceeds as a continuous, symmetry-breaking phase transition associated with the divergence of dielectric permittivity \cite{MartPT} and with disappearance of the polarization ${\bf{P}}_{\rm DW}$ within the domain wall interior \cite{StepJPCM}.
The internal domain wall polarization ${\bf{P}}_{\rm DW}$, existing only up to the Bloch-to-Ising transition point, is parallel or antiparallel to $[0\bar{1}1]$ direction. It is natural to ask whether the antiparallel ${\bf{P}}_{\rm DW}$ states could coexist within the same domain wall plane, similarly as the ferroelectric domains may coexist in the bulk of ferroelectric crystal \cite{EliseevYudin,winw,SaljeAPL2014,Sonin}.


In our previous work \cite{StepJPCM} we have studied the case when the ${\bf{P}}_{\rm DW}$ vector has an opposite orientation within neighboring domain walls, as indicated in Fig.\,\ref{fig1}\,b.
Such domain structure contains only the domain walls of the same helicity and it is thus a chiral structure as a whole.
 In contrast, in the present work we have explored a racemic domain structure, which contains an equal amount of  left-handed and right-handed areas on average as well as within each domain wall plane itself. This arrangement, depicted in the Fig.\,\ref{fig1}\,c comprises an interesting topological defect -- a borderline between domain wall regions of opposite helicity.
   This defect could be in general denoted as a polarization disclination line, recalling for example the analogy to the dislocation lines of crystal lattices and disclination lines of liquid crystal textures \cite{disclination}. However, for the reasons given below, we prefer to denote the specific defect studied here as an {\it Ising line}.
    The aim of the present paper is to describe the basic properties of this so far unexplored structural object.

 In order to learn about the structure of the Ising line, we have carried out standard phase-field simulations  based on numerical solution of the time-dependent Ginzburg-Landau equation for a phenomenological Ginzburg-Landau-Devonshire  energy functional complemented by exact calculation of the long-range electrostatic energy associated with the inhomogeneous profile of the polarization field \cite{simul}. The local mechanical equilibrium and strain compatibility is implicitly imposed by the method \cite{simul}. Further details about the phase-field approach in ferroelectrics can be found in Refs.\,\onlinecite{Ondrejkovic2009, HuChen3D, Khacha, Nambu, Slutsker2008}.

 In the results displayed here, we employed the same Ginzburg-Landau-Devonshire model for BaTiO$_3$ as in Ref.\,\onlinecite{StepJPCM}.
   The adjustable temperature parameter of the model was set to  118\,K, but similar results were obtained at other temperatures within the stability range of the rhombohedral ferroelectric phase as well.
Periodic boundary conditions were applied to both polarization and strain fields. Furthermore, the overall stress-free conditions were assumed (stress averaged over the periodic supercell is vanishing).
 The calculation was typically conducted at a $128\times128\times128$ sized discrete mesh with 0.5\,nm or 0.25\,nm spatial steps. This allows to inspect local polarization vectors at the scale of the perovskite elementary unit cell parameter.

     The initial configuration was close to the final relaxed configuration  shown  in Fig.\,\ref{fig1}\,c. A careful design of the initial configuration was an essential condition for keeping the Ising lines in the relaxed domain structure, because these lines are  highly mobile and easily annihilate. The architecture of our initial configuration obviously respected the basic requirements such as that the energetically very costly head-to-head or tail-to-tail polarization configurations should be avoided. Among others, this implies that the borderlines between domain wall regions with opposite ${\bf{P}}_{\rm DW}$ are all parallel to ${\bf{P}}_{\rm DW}$, see Fig.\,\ref{fig1}\,d.

Subsequently, the initial polarization field configuration has been fully relaxed according to the time-dependent Ginzburg-Landau equation \cite{simul, Ondrejkovic2009, StepJPCM}.
The final relaxed domain structure state indeed kept the overall architecture of the engineered initial state (Fig.\,\ref{fig1}\,c), at least as long as the distances between the  Ising lines were larger than about 3\,nm and the distance between the domain walls was larger than about 6\,nm.

\begin{center}
\begin{figure}
\includegraphics[width=5.5cm]{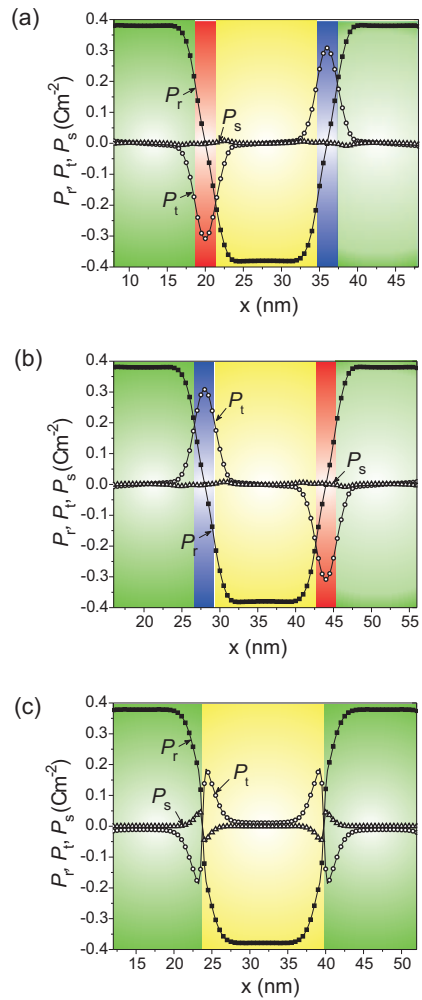}
\caption{Profiles of $P_{\rm r}, P_{\rm t}$, and $ P_{\rm s}$ polarization components along the $[100]$ direction as obtained from phase-field simulations for rhombohedral BaTiO$_3$ at 118\,K using the model described in the text. Data are taken from the front face of the simulation box (a) at the trajectory "1"  shown in Fig.\,\ref{fig1}c, (b) at the trajectory "2"  shown in Fig.\,\ref{fig1}c, and (c) at the intermediate trajectory passing through the Ising lines.
} \label{fig2}
\end{figure}
\end{center}

The local arrangement of the local polarization around the core of the Ising lines is shown in Fig.\,\ref{fig1}\,f.
For a more quantitative analysis of the resulting polarization profiles, it is useful to employ the symmetry-adapted  coordinates associated with the orthogonal unit vectors ${\bf{r}} \parallel [111]$
${\bf{s}} \parallel  [\bar{2}11]$, and ${\bf{t}} \parallel [0\bar{1}1]$, see Fig.\,\ref{fig1}\,c,e.
For example, the Bloch-like nature of the domain walls in the relaxed structure becomes apparent when
the  $P_{\rm r},P_{\rm t}, P_{\rm s}$ polarization components are traced across the domain wall (see Fig.\,\ref{fig2}\,a).
In the heart of the domain wall, where the bulk spontaneous polarization changes between  $ [111]$  and $[\bar{1}\bar{1}\bar{1}]$ orientations, the principal polarization component
$P_{\rm r}$ passes through zero.
However, at the same time, the additional polarization component
$P_{\rm t}$ appears, so that ${\bf{P}}_{\rm DW}$, the total polarization at $P_{\rm r}=0$ point, become nonzero.
 The exchange between $P_{\rm r}$ and $P_{\rm t}$ components on the path across the domain wall corresponds to the rotation of the polarization vector within the plane parallel to the domain wall. This is  similar to the case of ferromagnetic Bloch domain walls \cite{Kittel,StepJPCM}.
 In the domain walls shown in Fig.\,\ref{fig2}\,b, the polarization rotates in the same sense as the windings on a right-handed screw go; the opposite sense of rotation is encountered in domain walls depicted in Fig.\,\ref{fig2}\,a.

Fig.\,\ref{fig2}\,c displays the polarization profile along a trajectory that connects the antiparallel bulk domain states through the Ising line.
It demonstrates that within the Ising line, the polarization vector vanishes. In other words, this profile actually reminds of an Ising wall, and this is why the line observed here is denoted as  Ising line.  This strongly contrasts with analogical line defects known from ferromagnetism, so-called {\it N\'{e}el lines}, where the magnetization does {\it not vanish} in the domain wall center (instead, it only tilts out of the plane of the domain wall) \cite{disclination}.
 In this sense, the Ising lines obtained here are the lines of true singularities of the polarization distribution and in principle, the direction of the polarization is not defined in the core of the line at all.
The reason for this difference is obviously the importance of the electrostatics; ferroelectric
Neel line would be strongly penalized by the associated local depolarization field \cite{Levanyuk}. However,
 this penalty is less prohibitive than that of the 2D Neel wall, and in principle,  the
ferroelectric Neel line may also exist in some ferroelectric materials.

\begin{center}
\begin{figure}
\includegraphics[width=5.5cm]{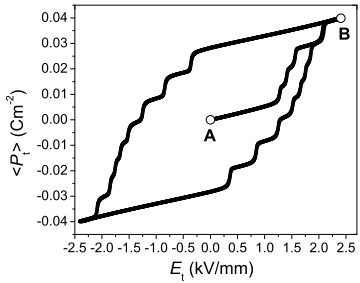}
\caption{Quasistatic hysteresis loop showing partial switching of the spontaneous domain wall polarization ${\bf{P}}_{\rm DW}$ by  the electric field applied along ${\bf t}\parallel [0\bar{1}1]$ direction. Results were obtained from phase-field simulations at 118\,K using 128x128x128 simulation box with a spatial step 0.5\,nm. The polarization profiles taken at the points marked  A and B are shown in Fig.\,4.} \label{fig3}
\end{figure}
\end{center}

We have also explored the stability of the domain structure with respect to applied bias electric fields. The width of the Ising line is about 2\,nm and such a broad thickness ensures its high mobility within the domain wall. Concerted drift of Ising lines can be forced by a homogenous bias field applied along  $\bf{t} \parallel [0\bar{1}1]$ direction. For example, in the simulations with 0.5\,nm step  the coercive field sufficient to remove all the Ising lines at 118\,K was only about 3\,kV/mm.

\begin{center}
\begin{figure}
\includegraphics[width=5.5cm]{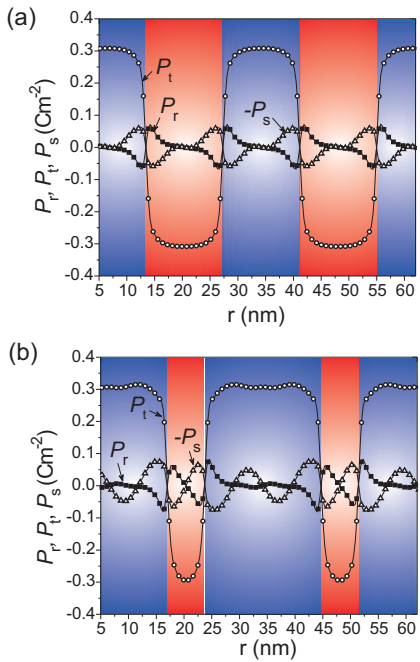}
\caption{Profiles of $P_{\rm r}, P_{\rm t}$ and $P_{\rm s}$ polarization components along ${\bf r}\parallel [111] $ direction, passing through the Ising line singularities. Panel (a) and (b) correspond to the points marked  A and B  in Fig.\,3 and represent the virgin state and partially switched state, respectively.
} \label{fig4}
\end{figure}
\end{center}

A hysteresis loop showing the variations of the ${{P}}_{\rm t}$ polarization component averaged over the whole supercell under a  periodically oscillating electric field is shown in Fig.\,\ref{fig3}. This hysteresis is related to the growth of the area with a favorably oriented ${{P}}_{\rm t}$ component, mediated by the motion of the Ising lines. The comparison of the virgin and partially switched polarization profiles  within the domain wall plane is shown in Fig.\,\ref{fig4}. The profile of ${{P}}_{\rm t}$ component  clarifies that the Ising line can  move as a soliton, without any essential change in its internal form.

Fig.\,\ref{fig4} also shows slight spacial oscillations of  the profiles of ${{P}}_{\rm s}$ and ${{P}}_{\rm r}$  components.
They are partially related to a slight bending of the domain wall plane around the Ising line (the wall is not completely planar).
On the other hand, the polarization profiles of the domain wall in the flat regions distant from the Ising line  are very similar to the ideal polarization profiles obtained in Refs.\,\onlinecite{StepJPCM,EliseevYudin}. In spite of a strong electrostatic interaction term in the model, domain wall profiles in these flat regions (Fig.\,\ref{fig2}ab) show  a clear indication of the characteristic double-kink in the normal (N\'{e}el) component of the polarization (${{P}}_{\rm s}$). It is worth noting that in some high-symmetry domains walls, \cite{Gu2014,Li2014,Yudin2012} such double kinks on ${{P}}_{\rm s}$ or ${{P}}_{\rm t}$ appear only after considering flexoelectric \cite{YudinNANOTECH2013} coupling terms, while here, in the case of the $[\bar{2}11]$-oriented Bloch (bistable \cite{Yudin2013}) wall, the flexoelectric coupling has actually only a negligible influence on the domain wall profile \cite{StepJPCM,EliseevYudin}. In either case,  the amplitude of these double-kinks (about one percent of the spontaneous polarization) is  much smaller than the magnitude of the ${{P}}_{\rm s}$  oscillations of Fig.\,\ref{fig4}, introduced by the presence of Ising lines themselves.

As already mentioned,  the Ising lines are parallel to the ${\bf{t}} \parallel [0\bar{1}1]$ direction, and in fact, in our simulation the polarization happens to be strictly constant along ${\bf{t}}$  direction. Therefore, it is convenient to inspect section of the simulated domain structure by a plane perpendicular to ${\bf{t}} \parallel [0\bar{1}1]$ direction, highlighted in Fig.\,5a. Polarization distribution in this plane (Fig.\,5b)  shows that the Ising lines in the  domain structure investigated here are actually forming a slightly distorted triangular lattice with mutual distances of about 14\,nm.

If we follow the polarization round an oriented circuit far from the core of given Ising line  (see Fig.\,5c), it is found that the polarization rotates once around the ${\bf{s}}$ direction \cite{Mermin}. Therefore,
the Ising line can be considered as an {\it edge} disclination of a unit strength \cite{disclination}. The nearest neighbor Ising lines within the same domain wall have always the opposite sense of the polarization rotation.
It is worth to note that in case of the flux closure vortexes observed in ferroelectric  thin films and ferroelectric nanodots, a similar construction would result in rotation around the core of the vortex and would be thus analogical to a {\it screw} disclination of a unit strength.

 The macroscopic (average) symmetry of the investigated domain structure is mmm, with principal axes along the ${\bf{r}} $, ${\bf{s}} $, and ${\bf{t}} $ directions. However, all the true mirror planes have fractional translation parts. There is no simple mirror symmetry plane passing through the individual Ising line, which is clearly a chiral object. The Ising line is translationally invariant along the ${\bf{t}} $ direction. In the continuum limit, the core of the Ising line is a line of zero polarization. The Jacobian matrix evaluated there has two complex conjugate eigenvalues with a positive real part and one zero eigenvalue. This implies that it is a line of singular points of saddle-node bifurcation type \cite{saddle-node}.

\begin{center}
\begin{figure}
\includegraphics[width=7.5cm]{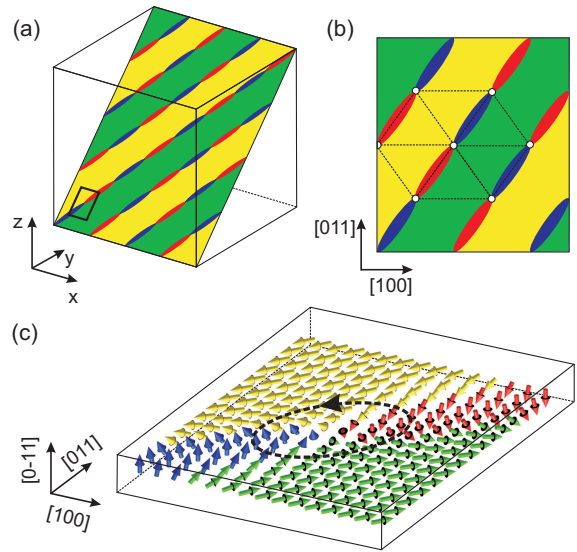}
\caption{ Calculated polarization field within the  plane perpendicular to the Ising lines (a) and its smaller part with indicated triangular lattice of the Ising lines (b). Panel (c) displays vicinity of one Ising line highlighted in (b) and an oriented closed circuit around its core.
} \label{fig5}
\end{figure}
\end{center}

In summary, we clarified the stability and nature of an interesting type of line defect expected to exist in ferroelectric Bloch walls as
a borderline between areas of opposite helicity.  It is found that this defect is highly mobile and its motion mediates the switching of the Bloch domain wall polarization and its helicity.
It was found that in contrast to Bloch lines and N\'{e}el lines, which play a similar role
in ferromagnets, the core of the ferroelectric defect revealed here is unpolarized. In this sense, it is appropriate to term this defect as Ising line.

We have proposed a simple ferroelectric domain structure with a triangular lattice of such Ising lines. This structure was demonstrated to be stable within the established  Ginzburg-Landau-Devonshire model for rhombohedral phase of BaTiO$_3$. It was found that
the internal structure of the Ising line can be described as a line of saddle-node type singular points, forming a polarization disclination of a unit strength.
 In a more simple words, this Ising line is a straight and about 2\,nm thick paraelectric nanorod of intrinsic nature, embedded within a polar but uncharged 180-degree  Bloch ferroelectric wall and  extended along a $\{011\}$ type direction.

The polarization vector within a circuit drown around the axis of the Ising line studies here  makes a full turn
 around the Bloch wall normal ${\bf s}$,  so that the defect structure  differs from more frequent cases where the polarization rotates around the axis of the defect, as for example in 2D skyrmions or the ferroelectric flux-closure vortexes. Similar Ising lines are obviously expected to exists in other kinds of ferroelectric Bloch walls.

\begin{acknowledgments}
This work is supported by the Czech Science Foundation
(Project CSF 15-04121S).
\end{acknowledgments}

\end{document}